\documentclass[twocolumn,preprintnumbers,amsmath,amssymb]{revtex4}
\usepackage{graphicx,epsfig}
\usepackage{dcolumn}
\usepackage{bm}
\begin{document}

\title{Anomalous Temperature Dependence of Heat Capacity During a 
Cooling-Heating Cycle in Glassy Systems}
\author{Dwaipayan Chakrabarti and Biman Bagchi\footnote[2] 
{For correspondence: bbagchi@sscu.iisc.ernet.in}} 
\affiliation{Solid State and Structural Chemistry Unit, Indian Institute of 
Science, Bangalore 560012, India}

\date{\today}

\begin{abstract}

Anomalous temperature dependence of heat capacity of glassy systems during a 
cooling-heating cycle has remained an ill-understood problem for a long time.
Most of the features observed in the experimental measurement of the heat 
capacity of a supercooled liquid are shown here to be adequately explained by 
a general model. The model that we propose is motivated by the success of 
landscape paradigm, and describes $\beta$ relaxation in terms of a collection 
of two-level systems and conceives $\alpha$ relaxation as a $\beta$ relaxation 
mediated cooperative transition in a double-well. The anomalous sharp rise in 
the heat capacity observed during heating is shown to have a kinetic origin, 
being caused by delayed energy relaxation due to nonequilibrium effects.

\end{abstract}

\maketitle 

The glass transition region is characterized by both thermodynamic and kinetic
anomalies \cite{Angell-Ngai-McKenna-McMillan-Martin, Ediger-Angell-Nagel, 
Debenedetti-Stillinger}. One experimentally finds a sharp rise in the measured 
heat capacity of a liquid during rate heating which follows prior cooling at a 
constant rate, the cycle being well extended on either sides of the glass 
transition region \cite{Moynihan-JPC-1974, Moynihan-JACeS-1976, 
DeBolt-JACeS-1976}. The overshoot of the heat capacity is taken to be the 
signature of a glass to liquid transition. While the details of the magnitude of 
the measured heat capacity vary with the cooling rate $q_{c}$ and the heating 
rate $q_{h}$, the general features remain qualitatively the same over a rather 
wide range of $q_{c}$ and $q_{h}$. In the glass transition region, one also 
encounters with highly nonexponential relaxation of enthalpy, stress and 
polarization, which is often described by the Kohlrausch-Williams-Watts (KWW) 
form, and a very rapid increase of the shear viscosity of the liquid over a 
narrow temperature range \cite{Angell-Ngai-McKenna-McMillan-Martin, 
Ediger-Angell-Nagel, Debenedetti-Stillinger}. Satisfactory explanation of these 
anomalies has eluded us for a long time. The glassy dynamics is often 
considered to be a manifestation of an underlying phase transition 
\cite{Gibbs-DiMarzio-JCP-1958, Wolynes-PNAS-2000}. However, no consensus has 
still been reached as to the thermodynamic versus kinetic origin of the observed 
anomalies. Another well-known method to study dynamics in supercooled liquids 
and glasses is to measure the frequency dependence of heat capacity 
$C_{v}(\omega)$, where the relaxation time of the heat capacity is the energy 
relaxation time. As discussed by Birge and Nagel, and by Oxtoby, the frequency 
dependent heat capacity could provide the much needed connection between the 
thermodynamical and kinetic anomalies \cite{Birge-Nagel-PRL-1985, 
Oxtoby-JCP-1986}. Oxtoby has, in fact, derived an elegant relationship between 
the time dependence of heat capacity and the time dependence of the fictive 
temperature \cite{ Oxtoby-JCP-1986}. 

It is a common practice to characterize the nonequilibrium state of the liquid 
encountered in time domain experiments by the fictive temperature $T_{f}$, 
which, as defined by Tool and Eichlin \cite{Tool-Eichlin}, is the temperature at 
which the nonequilibrium value of a macroscopic property would equal the 
equilibrium one. If cooling is continued through the supercooled regime, the 
structural relaxation eventually becomes too slow to be detected on the 
experimental time scale, resulting in a limiting fictive temperature. The 
limiting fictive temperature $T^{L}_{f}$ obtained upon cooling is known to 
depend on $q_{c}$, while the glass transition temperature $T_{g}$, as measured 
experimentally from a cooling-heating cycle, is dependent on both $q_{c}$ and 
$q_{h}$, a shift to higher values being observed for faster rates. If the rates 
of cooling and heating are taken to be the same, that is $ q_{c} = q_{h} = q $, 
the dependence of $T_{g}$ on $q$, as shown elegantly by Moynihan {\it et al.} 
\cite{Moynihan-JPC-1974}, is given by 
\begin{equation}
\frac{dlnq}{d(1/T_{g})} = - \Delta h ^{*} / R,
\end{equation}
where $\Delta h ^{*}$ can be interpreted as the activation enthalpy for the 
structural relaxation in effect and $R$ is the universal gas constant. As 
pointed out by Moynihan {\it et al.}, it is important for the validity of the 
above relationship that the material be cooled and reheated not only at the same 
rate but the cycle be extended well beyond the glass transition range. 
$T^{L}_{f}$ is also shown to have an identical dependence on $q_{c}$ 
\cite{Moynihan-JACeS-1976}, which has recently been reproduced in some model 
glassy systems \cite{Halpern-Bisquert-JCP-2001}. 

The theoretical analysis of nonequilibrium heat capacity is a non-trivial 
problem and has been addressed in great detail by Brawer \cite{Brawer-1983, 
Brawer-JCP-1984}, J\"{a}ckle \cite{Jackle}, and in recent time by Odagaki and 
coworkers \cite{Odagaki-PRE-2001-2002}. The two widely used expressions of 
equilibrium heat capacity at constant volume are given by 
\begin{equation}
C_{v}(T) = \frac{\partial E(T)}{\partial T}
\end{equation}
and
\begin{equation}
C_{v}(T) = \frac{<(\Delta E(T))^2>}{k_{B}T^2},
\end{equation}
where $ <(\Delta E(T))^2> $ is the mean square energy fluctuation at temperature
$T$. As is well known, these two are equal in equilibrium ergodic system. 
However, they need not be equal in nonequilibrium system.  

In this Letter, we present a theoretical analysis of heat capacity of a general 
model of glassy relaxation and show that essentially {\it all} the features 
observed during a cooling-heating cycle can be explained satisfactorily. The 
model has been conceived in the spirit of the energy landscape concept 
\cite{Stillinger-Science-1995, Sastry-Debenedetti-Stillinger-Nature-1998}, where 
one describes the system as an ensemble of double-well potentials with a broad 
distribution of barrier heights and asymmetries between the two minima for the 
local structural rearrangements \cite{Pollak-Pike-PRL, Gilroy-Phillips, 
Buchenau_PRB}. The model, based on the framework of $\beta$ 
organized $\alpha$ process, envisages a $\beta$ process as a transition in a 
two-level system (TLS). The waiting time before a transition can occur from the 
level $i$, labeled either $0$ or $1$, is taken to be random, and is drawn from 
the Poissonian probability density function given by
\begin{equation} 
\psi_{i}(t) = \frac{1}{\tau_{i}} exp(-t/\tau_{i}), ~~~~~~~~~~~~~~i = 0, 1,
\end{equation}
where $\tau_{i}$ is the average time of stay in the level $ i $. If $p_{i}$ 
denotes the canonical equilibrium probability of the level $i$ being occupied, 
detailed balance gives the following relation
\begin{equation}
K(T) = \frac{p_{1}(T)}{p_{0}(T)} = \frac{\tau_{1}(T)}{\tau_{0}(T)} 
= exp[-\epsilon/(k_{B}T)],
\end{equation}
where $ K $ is the equilibrium constant for the two levels $0$ and $1$, which 
are taken to have energies zero and $\epsilon$, respectively, and $k_{B}$ is 
the Boltzmann constant. In our model, an $\alpha$ process is conceived as a 
cooperative transition from one well to another in a double-well, subject to 
the establishment of a certain condition. 
\begin{figure}[tb]
\epsfig{file=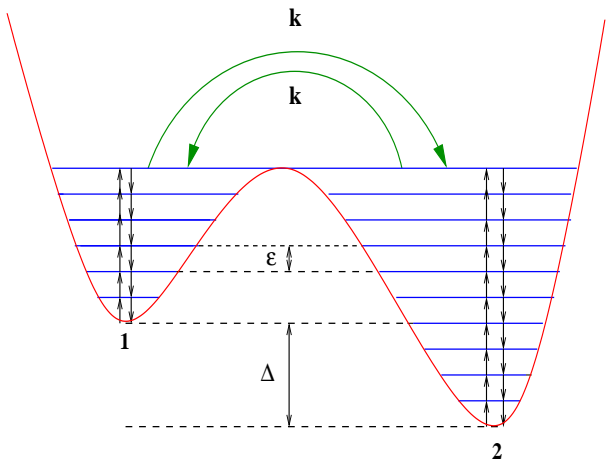,height=6cm,width=8cm,angle=0}
\caption{A schematic representation of the model under consideration. The 
horizontal lines within a well represent different excitation levels. Note that 
the energy levels are in general degenerate, as they correspond to the sum of 
the energies of individual TLSs in the collection.}
\end{figure}
See Fig. 1 for a pictorial representation of the model. Each of the two wells, 
labeled $1$ and $2$, comprises a collection of $N_{i}~(i = 1$ and $2$, 
respectively $)$ identical, non-interacting TLSs of such kind. For a collection 
of $N_{i}~ (i = 1, 2)$ TLSs, a variable 
$\zeta_{j}^{i}(t),~(j = 1, 2, ....., N_{i})$ is defined, which takes on a value 
$0$ if at the given instant of time $t$ the level $0$ of the TLS $j$ is occupied 
and $1$ if otherwise. $\zeta_{j}^{i}(t)$ is thus an occupation variable. The 
variable $Q_{i}(t)~(i = 1, 2)$ is then defined as
\begin{equation}
Q_{i}(t) = \displaystyle \sum_{j=1}^{N_{i}} \zeta_{j}^{i}(t).
\end{equation}
$Q_{i}(t)$, which serves to describe the level of instantaneous excitation in a 
collection of TLSs, is therefore a stochastic variable in the discrete integer 
space $[0, N_{i}]$. {\it An $\alpha$ process occurs only when all the $\beta$ 
processes (TLSs) in a well are simultaneously excited, i.e., $Q_{i} = N_{i}$.}
There is a finite rate of transition $k$ from either wells when this condition 
is satisfied. Within the general framework of the model, the double-well becomes 
asymmetric when $ N_{1} \neq N_{2} $.

Note that the double-well model has been widely used to describe relaxation in 
glassy liquids and we do not specify the microscopic nature of the two states
any further. The main new ingredient we introduce is the condition that 
transition between the two wells is controlled by the collective variable 
$Q_{i}(t)$. Note also that the dynamics embodied in Fig. 1 is not the standard 
chemical dynamics in a bistable potential because the microscopic energy levels 
are in general degenerate. One should also note that in the energy landscape 
picture the transition from the initial state can occur to a limited number 
(but greater than unity) of final states.
 
In Fig. 2, we show the heat capacity versus temperature $C-T$ curve obtained for 
our model. The curve looks very close to the ones observed in experiments.
\begin{figure}[tb]
\epsfig{file=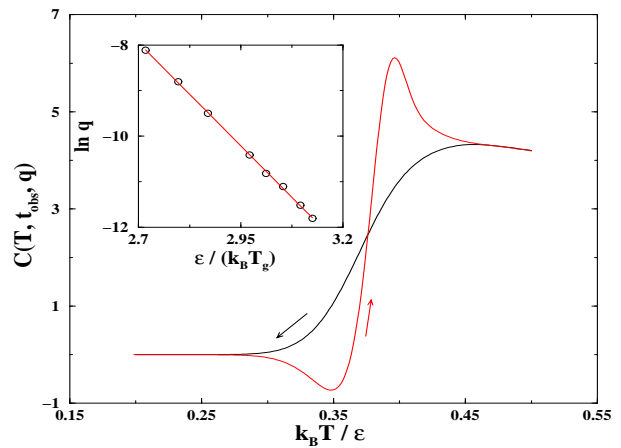,height=6cm,width=8cm,angle=0}
\caption{The heat capacity versus reduced temperature plot for the system when 
subjected to a cooling-heating cycle with $q = 7.5 \times 10^{-5}$ in reduced 
units. The inset shows the plot of the logarithm of the cooling rate $q$ versus 
the reciprocal of the $T_{g}$. The slope of the linear fit to the data equals 
$-9.04$ in appropriate temperature units.}
\end{figure}
{\it Note the sharp rise in heat capacity during heating.} We now briefly 
describe the calculation procedure. The probability $P_{i}(n;T,t)$ that the 
stochastic variable $Q_{i}$ takes on a value $n$ at temperature T and time $t$ 
can be shown to satisfy the following stochastic master equation 
\cite{van Kampen}:
\begin{eqnarray}
\frac {\partial P_{i}(n;T,t)}{\partial t} 
&=& [(N_{i} - n + 1)/\tau_{0}(T)]P_{i}(n - 1;T,t) \nonumber \\ 
&+& [(n + 1)/\tau_{1}(T)]P_{i}(n + 1;T,t) \nonumber \\ 
&-& [(N_{i} - n)/\tau_{0}(T)]P_{i}(n;T,t) \nonumber \\ 
&-& (n/\tau_{1}(T))P_{i}(n;T,t) - k~\delta_{n,N_{i}}~P_{i}(n;T,t) \nonumber \\ 
&+& k~\delta_{n,N_{i\pm1}}~\delta_{j,i\pm1}~P_{j}(n;T,t),
\end{eqnarray}
where the '$+$' and '$-$' signs in the indices of the Kronecker delta are for 
$i = 1$ and $2$, respectively. 
The total energy of the system at time $t$ can therefore be given by  
\begin{equation}
E(T, t) = \displaystyle \sum_{n=0}^{N_{1}} P_{1}(n;T,t)~(N_{2} - N_{1} + n )~
\epsilon + \displaystyle \sum_{n=0}^{N_{2}} P_{2}(n;T,t)~n~\epsilon,
\end{equation}
where the lowest level of the well $2$ is taken to have zero energy. The set of 
equations, given by Eq. (7) for $ n = 0, 1, ..., N_{i} $ and $i = 1, 2$, is 
solved numerically by the matrix method, where the solution is expanded in terms 
of the eigenvectors and the eigenvalues of the transition matrix, and the 
coefficients of the expansion are evaluated from the initial distribution. Once 
we know $P_{i}(n;T,t)$, we can calculate the heat capacity $C$, as discussed 
below, from an equation, which is essentially a form of Eq. (2) modified to 
incorporate the nonequilibrium effects.

The system, when subjected to cooling or heating at a constant rate, can be
envisaged to undergo a series of instantaneous temperature changes, each in 
discrete step of $\Delta T$ in the limit $\Delta T \rightarrow 0$, at time 
intervals of length $\Delta t$, whence $ q = \Delta T / \Delta t $ 
\cite{Moynihan-JPC-1974}. A pictorial representation of the temperature control 
during a cooling process with finite $\Delta T$ was given by 
J\"{a}ckle \cite{Jackle}. If we consider a time interval at the beginning of 
which the temperature has been changed from $T$ to $T^{\prime} = T + \Delta T$, 
the waiting time $t_{obs}$ before an observation can be made is restricted by 
$\Delta t$. The heat capacity $C$, measured at a time $t_{obs}$ subsequent to a 
temperature change from $T$ to $T^{\prime} = T + \Delta T$, is not stationary in 
time unless $t_{obs}$ is long enough for the equilibrium to be established. The 
measured heat capacity ( as is the energy ) then becomes a function of the rate 
of cooling / heating as well, apart from $T$ and $t_{obs}$. The dependence of 
$C$ on $q_{c}$ / $q_{h}$ implies that the measured heat capacity of a 
nonequilibrium state depends on the history of the preparation of that state. 
Here we restrict ourselves to the case, where $q_{c} = q_{h} = q$. We therefore 
calculate $C(T, t_{obs}, q)$ from the following equation:
\begin{equation}
C(T, t_{obs}, q) = \lim_{\Delta T \rightarrow 0} 
\frac{E(T + \Delta T, t_{obs}, q) - E(T, 0, q)}{\Delta T},   
\end{equation}
where the energies can be obtained from Eq. (8). In the present calculation, we 
have taken $t_{obs} = \Delta t$. Throughout the cycle the transition rates are 
assumed to be tuned with the heat bath temperature $T$.

Fig. 2 is the result of the model calculation where $N_{1} = 6$ and 
$N_{2} = 10$. Temperature $T$ is throughout expressed in reduced units of 
$\epsilon / k_{B}$ with $\epsilon$ taken to be unity. We have set 
$\Delta T = \pm 0.0015 $ in reduced units. The correspondence to real units is 
discussed later. The model assumes the presence of an energy barrier for all the 
intra-well transitions. If $\epsilon^{\ddagger}_{i}$ be the energy barrier to 
the transition from the level $i$ in a TLS, the transition state theory (TST) 
allows one to write 
$\tau_{i}(T) = h / (k_{B}T) exp[\epsilon^{\ddagger}_{i}/(k_{B}T)]$, where $h$ is 
the Planck constant. Here we have set $\epsilon^{\ddagger}_{1} = 8~\epsilon$. We 
express time also in reduced units, being scaled by $\tau_{1}(T_{h})$, where 
$T_{h}$ is the highest temperature considered in a cooling-heating cycle. Note 
that the cycle starts with the equilibrium population distribution at $T_{h}$. 
The inter-well transition rates are equal and independent of temperature. We 
have taken $k^{-1} = 0.50 $ in reduced time units. In order to further explore 
the merit of the model in reproducing the experimental results, we have 
investigated the cooling / heating rate dependence of $T_{g}$ for our model. The 
latter has been taken as the temperature of onset of the heat capacity increase 
as observed during heating \cite{Angell-Science-1995}. The $ln q$ versus $1/T_{g}$ 
plot, as shown in the inset of Fig. 2, is linear with negative slope in complete 
accordance with the experimental observations. The slope gives a measure of the 
energy of activation for the relaxation being in operation.

\begin{figure}[tb]
\epsfig{file=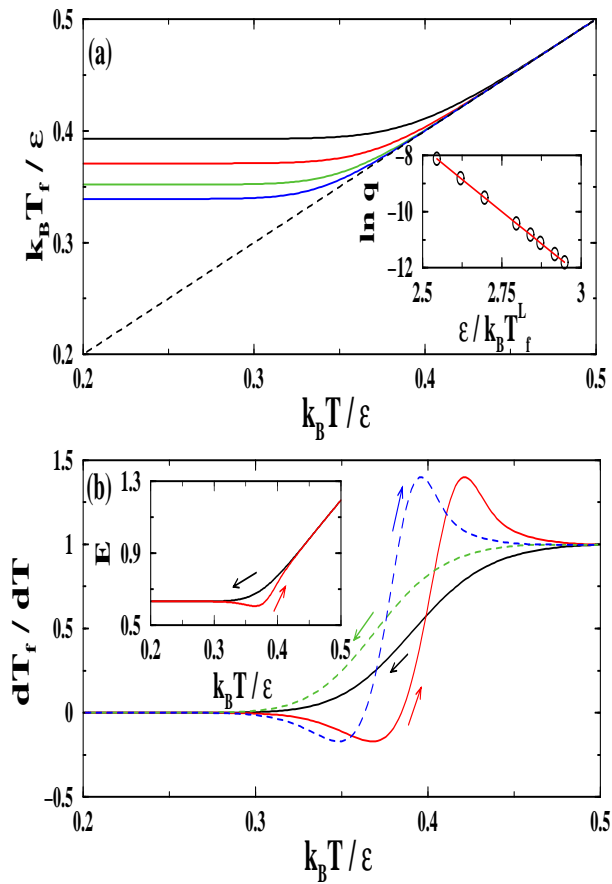,height=12cm,width=8cm,angle=0}
\caption{(a) Plot of the fictive temperature $T_{f}$ versus the heat bath 
temperature $T$ in reduced units for different cooling rates: 
$q = 3.0 \times 10^{-4}, 7.5 \times 10^{-5}, 2.0 \times 10^{-5}, 
7.5 \times 10^{-6}$ from top to bottom. The dashed line traces the $T_{f} = T$ 
line. The inset shows the dependence of the limiting fictive temperature 
$T^{L}_{f}$ obtained upon cooling on the rate of cooling. The slope of the 
linear fit to the data is $-9.13$ in appropriate temperature units.
(b) The $dT_{f}/dT$ versus reduced temperature plot for the system with 
$N_{1} = 6$ and $N_{2} = 10$ when subjected to a cooling-heating cycles. The 
solid line is for $q = 3 \times 10^{-4}$, and the dashed line is for 
$q = 7.5 \times 10^{-5}$, both in reduced units. The inset shows the evolution 
of the energy of the system during a cooling-heating cycle with 
$q = 7.5 \times 10^{-5}$ in reduced units.}   
\end{figure}

Fig. 3(a) shows a $T_{f}$ versus $T$ plot for different cooling
rates, where the fictive temperature is calculated in terms of energy. The 
freezing of structural relaxation within the experimental time scale is evident 
from the attainment of a limiting fictive temperature. In the inset of 
Fig. 3(a), we show the plot of the logarithm of the cooling rate $q$ versus the
reciprocal of the limiting fictive temperature $T^{L}_{f}$ obtained on cooling. 
The linearity of the plot with negative slope is  again in excellent agreement 
with the experimental results. A plot of $dT_{f}/dT$ versus $T$ is displayed 
in Fig. 3(b) for two different cooling / heating rates. This is also in accord 
with the experimental observation.

So far we have presented the results in reduced units. It would be useful to 
have, at this point, an estimate of the real terms for reasonable input. For 
$\epsilon / k_{B} = 600~K$, the cooling and heating rates explored here range 
from $0.0085$ to $0.35~K s^{-1}$, while the temperature window we have looked 
into lies between $300~K$ and $120~K$. One should note that these rates are of 
the same order of magnitude as practised in experiments.    
     
In order to probe the origin of the behavior of the calculated heat 
capacity, and also of $dT_{f}/dT$, during heating, we plot the energy $E$ versus 
the heat bath temperature $T$ during a cycle, as shown in the inset of 
Fig. 3(b). The fictive temperature evolves in an identical fashion as the 
energy. Note that the energy or the fictive temperature goes down during the 
initial period of heating before it starts increasing. The reason is as follows. 
The presence of an energy barrier for all the intra-well transitions results in 
a slow down of the elementary relaxation rates as the system is subjected to 
rate cooling. The system eventually gets trapped into a nonequilibrium glassy 
state on continued cooling. As one subsequently starts heating, the rate of the 
elementary relaxation increases, thus causing a delayed energy relaxation. This 
explanation further gains support from the calculated heat capacity being 
negative \cite{Odagaki-PRE-2001-2002}. 

\begin{figure}[tb]
\epsfig{file=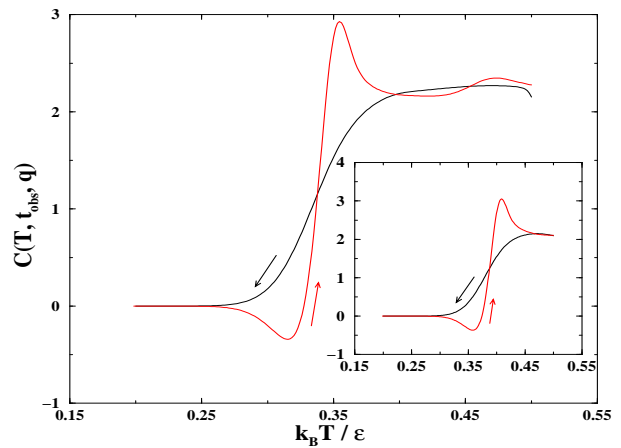,height=6cm,width=8cm,angle=0}
\caption{The heat capacity versus reduced temperature plot for the system with
$N_{1} = 3 $ and $ N_{2} = 5$ when subjected to a cooling-heating cycle 
with $q = 5 \times 10^{-6}$ in reduced units. The inset shows the same plot
for $q = 1.5 \times 10^{-4}$ in reduced units. The axis labels for the inset, 
being same as those of the main one, are not shown.}
\end{figure}

Fig. 4 shows a heat capacity versus temperature plot during a cooling-heating 
cycle for our proposed model obtained with a different set of parameters so 
chosen that the $\alpha$ relaxation becomes more probable within the observation 
time. A repeat of the non-monotonic pattern of the heat capacity, although in 
much smaller scale, is notable at higher temperatures during heating for slow 
cooling and heating. However, there has not been any report in the literature, 
to the best of our knowledge, of such an observation made experimentally. For 
faster rates, this repeat pattern vanishes as evident from the inset of Fig. 4. 
It is, therefore, reasonable to attribute this to the $\alpha$ relaxation.
  
A few comments regarding the present work are in order:\\
\noindent (1) The hysteresis in the $C$ versus $T$ plot, and also the 
overshoot of the heat capacity observed during heating, become progressively 
weaker as the cooling and heating rates decrease, and eventually vanish for 
sufficiently slow rates. This is again in agreement with the long known 
experimental results.\\
\noindent (2) While the elementary relaxation rates evolve with the heat bath 
temperature, {\it it is the slow population relaxation that gives rise to the 
nonequilibrium effects.}\\
\noindent (3) The relaxation of energy to its equilibrium value following a 
small temperature jump slows down rapidly as the temperature is lowered. This is
due to the activated dynamics assumed here for transition between the two levels 
in a TLS. From the temperature variation of this relaxation time, one can have 
an estimate of the temperature where the system starts falling out of the 
equilibrium while cooling at a given constant rate.\\
\noindent (4) Our model, while simple and microscopic, is quite general and 
also detailed. Its success in reproducing all the aspects of experimental 
results on heat capacity during the cooling-heating cycle is noteworthy. 
Another important aspect is that we need not invoke any singularity, 
thermodynamic or kinetic, to explain the anomalies. It is worth mentioning 
here that a similar model can describe many aspects of nonexponential 
relaxation observed in glassy liquids \cite{Chakrabarti-Bagchi}. We are 
currently investigating the relationship of the heat capacity anomaly with the 
fragility of the system \cite{Angell}.
 
We thank Mr. R. Murarka for several helpful discussions. This work was supported 
in parts by grants from CSIR and DST, India. DC acknowledges the University 
Grants Commission (UGC), India for providing the Research Fellowship.

\end{document}